\documentclass[]{pasj01}

\newcommand*{\mydprime}{^{\prime\prime}\mkern-1.2mu}

\begin{document} 
%\Received{}%{yyyy/mm/dd}
%\Accepted{}%{yyyy/mm/dd}
%\Published{yyyy/mm/dd}

\title{Searching For  the 380 GHz H$_{2}$O emission from the High-z lensed QSO MG J0414$+$0534}

%%% begin:list of authors
% Do NOT capitalize all letters in "textsc".
\author{Cheng-Yu \textsc{Kuo}\altaffilmark{1}%
\thanks{Present Address is cykuo@mail.nsysu.edu.tw}}
\altaffiltext{1}{Physics Department, National Sun Yat-Sen University,
No. 70, Lien-Hai Rd, Kaosiung City 80424, Taiwan, R.O.C }
\email{cykuo@mail.nsysu.edu.tw}

\author{Sherry H. \textsc{Suyu}\altaffilmark{2,3}}
\altaffiltext{2}{Max-Planck-Institut f\"{u}r Astrophysik, Karl-Schwarzschild-Str.~1, 85748 Garching, Germany}
\altaffiltext{3}{Physik-Department, Technische Universit\"at M\"unchen, James-Franck-Stra\ss{}e~1, 85748 Garching, Germany}
\email{suyu@mpa-garching.mpg.de}

\author{Violette \textsc{Impellizzeri}\altaffilmark{4,5}}
\altaffiltext{4}{Joint ALMA Office, Alonso de C$\acute{\rm o}$rdova 3107,
Vitacura, Santiago, Chile}
\email{Violette.Impellizzeri@alma.cl}

\author{James A. \textsc{Braatz}\altaffilmark{5}}
\altaffiltext{5}{National Radio Astronomy Observatory, 520 
Edgemont Road, Charlottesville, VA 22903, USA}
\email{jbraatz@nrao.edu}
%%% end:list of authors

%% `\KeyWords{}' always has to be placed before `\maketitle'.
 %Do NOT move this preamble from here!

\maketitle

\begin{abstract}
We report the result of our search for the 380 GHz H$_{2}$O line emissions from the quadruply lensed QSO MG J0104+0534 at z $=$ 2.639 with the Atacama Large Millimeter/submillimeter Array (ALMA). Our observation shows a tentative detection of  the 380 GHz line from the lensed QSO, and line spectrum shows a broad spectral distribution that has a velocity width of $\sim$290 km~s$^{-1}$ and a peak flux of $\sim$0.8 mJy. The integrated-intensity map of the H$_{2}$O line show lensed emissions at the A1 and A2 component of the QSO, with the A2 component slightly resolved. The integrated line flux ratio between the A1 and A2 component shows unexpected difference with the continuum flux ratio. Based on our gravitational lens modeling assuming our tentative detection is real, this flux ratio anomaly would suggest that the 380 GHz line emissions come from two or three spatially displaced locations in the QSO, with the dominant one located at the position of the continuum emission from the QSO and the other one(s) displaced from the continuum by $\sim$1.5 kpc on the source plane.
\end{abstract}

\section{Introduction}

Luminous 22 GHz H$_{2}$O maser emissions from circumnuclear environments (megamaser disks) in
active galaxies have been shown to provide unique tools to probe several important astrophysical questions (see reviews in Lo 2005).   
In addition to allowing the investigation of the sub-parsec/parsec scale gas distribution in active galactic nuclei (e.g. Greenhill 2003), megamaser disks that display Keplerian rotation, such as the one in the archetypal maser galaxy NGC 4258 (e.g. Herrnstein et al. 1999), further allow one to determining
the masses of supermassive BHs ($M_{\rm BH}$) to a few percent-level accuracy (e.g. Reid et al. 2009; Kuo et al. 2011; Gao et al. 2016; Zhao et al. 2018) and making direct precise measurements of angular-diameter distances beyond 100 Mpc in a single step (Kuo et al. 2013; Kuo et al. 2015; Gao et al. 2016), which in turn can lead to accurate Hubble constant determination (Braatz et al. 2018).    
        
Previous surveys of extragalactic 22 GHz maser emissions primarily targeted active galaxies in the local Universe (i.e. z $\lesssim$ 0.05). Nonetheless, out of the interest of studying properties of active galactic nuclei at high redshifts (i.e. z $\ge$ 1) using H$_{2}$O maser emission, several attempts have also been made to look for the 22 GHz masers from high-z gravitationally lensed galaxies (Impellizzeri et al. 2008; McKean et al. 2011; Castangia et al. 2011). 

In these high-z maser surveys, magnification provided by the gravitational lensing effect is expected to help boosting the maser fluxes and overcoming the sensitivity limitations that lead to non-detections in maser surveys at lower redshifts (e.g Bennert et al. 2009). Using the 100-m radio telescope in Effelsberg, Impellizzeri et al. (2008) made the first and the only discovery of the H$_{2}$O maser emission from galaxies with z $\ge$ 1 the gravitationally lensed QSO MG J0414+0532 (z $=$ 2.639). The strongest maser features from this source were detected at the $\sim$2-3 mJy level.

To see whether the maser emission originates from a circumnuclear disk at the center of the QSO, one of the co-authors (C. M. V. Impellizzeri) and her collaborators have attempted to image the maser distribution at milliarcsecond scale using the technique of Very Long Baseline Interferometry (VLBI), but no detection was made at the VLBI scale. A likely cause of the non-detection is that the fluctuation of the strong radio continuum (flux $\sim$560 mJy; Impellizzeri et al. 2008) of the QSO jet caused by instrumental instability overwhelms the weak line signals and prevents one to image maser emissions at milliarcsecond resolution. 

To overcome this difficulty, one promising approach is to image the H$_{2}$O maser emissions at the submillimeter wavelengths. Based on the theoretical predictions provided by Gray et al. (2016), it is shown that some of the submillimeter water maser lines can become significantly stronger than the 22 GHz maser emissions under certain physical conditions. Given the expectation that the synchrotron emission from the jet would be substantially weaker at submillimeter, the maser lines can thus stand out from the image that is affected by the instrumental instability.

In this work, we aim to search for the submillimeter water maser emissions from the lensed quasar MG J0414+0532 with the Atacama Large Millimeter/submillimeter Array (ALMA). We would like to see if we can detect emissions that are stronger than the 22 GHz masers for future VLBI followups.  In section 2, we explain why the 380 GHz transition is a promising maser line to search. The results of our ALMA observation will be shown in section 3. In section 4, we will explain the flux ratio anomaly seen in the integrated-intensity map of the 380 GHz line based on gravitational lens modeling and make predictions for future ALMA observations. The conclusion of this paper will be made in section 5.

\section{Observations and Data Reduction}
\subsection{Line Selection}
%To search for strong submillimter H$_{2}$O maser lines with ALMA towards MG J0414+0534, we primarily target maser transitions having rest frequencies $\nu_{0}$ above $\sim$300 GHz since all maser lines with $\nu_{0}$ $\lesssim$ 300 GHz from this source are redshifted outside all currently available ALMA bands. 
To search for strong submillimter water maser lines with ALMA towards MG J0414+0534, we primarily target maser transitions having rest frequencies above $\sim$300 GHz, a frequency below which all maser lines are redshifted outside all currently available ALMA bands. 

Among all possible submillimeter H$_{2}$O line candidates, we are most interested in the line having rest frequency $\nu_{0}$ of 380.19737 GHz (the 380 GHz line hereafter). Similar to the famous 22 GHz maser line, the 380 GHz line belongs to the maser transitions involving the \emph{back-bones levels}, which are expected to display large line strengths (e.g. de Jong 1973). Moreover, this line can be excited at similar gas density and temperature as for the 22 GHz maser. Based on the theoretical modeling (Gray et al. 2016), the strength of the 380 GHz maser line can be brighter than the 22 GHz one by a factor of 10$-$1000 (depending on the degree of maser saturation) if the molecular clouds responsible for maser emission can be heated to be above 1000 K. This is likely to occur because high-z quasars are expected to accrete close to the Eddington limit (Shen 2013) and the masing gas is thus exposed to stronger X-ray radiation which can heat the gas to be above 1000 K. If this maser line can be detected with high signal-to-noise ratio, then the follow-up imaging of the lensed maser emission with the Global mm VLBI Array (GMVA) would not only enable the first direct imaging of an accretion disk and dynamical BH mass measurements at z $>$ 2, but also provide a new cosmological probe to study dark energy.

In addition to the 380 GHz line, we also search for the line with $\nu_{0}$ of 336.22794 GHz (the 336 GHz line hereafter) as a by-product of this work because this transition can be accessible with the same spectral set-up when observing the 380 GHz line. The 336 GHz line belongs to the category of radiatively pumped masers which requires different physical conditions from those of collisionally pumped masers (e.g. the 22 GHz and 380 GHz masers). This transition can also be stronger than the 22 GHz maser under certain physical conditions.

\subsection{ALMA Observations}
The observations of MG J0414+0534 were conducted with ALMA at Band 3 with four execution blocks taken between 2017 December 13 and 2017 December 18 as part of Cycle 5 (Project ID : 2017.1.00316.S). The number of antenna involved in the observations is 45 in the first three executions and 44 in the last execution. The phase center was $\alpha$ = 04$^{\rm h}$14$^{\rm m}$37$^{\rm s}$.7699, $\delta$ = $+$05$^{\circ}$34$^{\prime}$42$\mydprime$.407 (ICRS). The total observing time including calibration and overheads is 3.96 hours.

The ALMA correlator was configured to four spectral windows (spw) with band centers at 92.396, 93.977 GHz, 104.479 GHz, and 106.081 GHz (spw 27, 23, 25, 19), respectively. Spw 27 and 25 have a bandwidth of 0.938 GHz with 1920 channels within each band. These two windows were primarily used to search for the 336 GHz and the 380 GHz lines, which are redshifted to 92.39570 GHz and 104.47830 GHz, respectively. Given the assumption that the 380 GHz line has the same linewidth as the 22 GHz maser seen in Impellizzeri et al. (2008) and Castangia et al. (2011), the $\sim$1 GHz bandwidths of spw 25 and spw 27 were expected to be sufficiently broad to well cover the spectral distribution of the maser emissions.  

Spw 23 and 19 have a bandwidth of 2 GHz with 128 channels in each band. We use these two bands to image the continuum emissions of the lensed QSO in order to get precise continuum positions of the QSO for gravitational lens modeling.

\subsection{Data Reduction}
We performed data reduction of the ALMA data with the Common Astronomy Software Application Package (CASA; McMullin et al. 2007). After the initial data flagging, the amplitude and bandpass calibrations of the visibility data were carried out with the calibrator J0423$-$0120. The phase calibration was performed using the calibrator J0427$+$0457. To have a robust detection of the maser lines and precise measurements of continuum fluxes and QSO positions, we performed self-calibration (phase-only) on the visibility data for all four spectral windows to remove the residual phase variations (up to $\sim$40$^{\circ}$) on timescales as short as 1$\sim$5 minutes. After self-calibration, we conduct continuum and spectral-line imaging with the CASA task $TCLEAN$. The continuum imaging was carried out using the uniform weighting in order to achieve the highest resolution for precise position measurements. The resulting synthesis beam is 0$\mydprime$.35$\times$0$\mydprime$.22, with PA measured in degrees east of north of $-$67$^{\circ}$. The 1$\sigma$ noise of the continuum image made from averaging across a spectral width of 1.78 GHz is 0.036 mJy. 

To conduct the spectral-line imaging with the visibility data taken in spw 25 and 27, we first subtract the continuum emission in the visibility domain with the CASA task $UVCONTSUB$. To identify the baseline channels for continuum subtraction in spw 25, we first assume that the 380 GHz line has the same optical velocity (with respect to the recession velocity of the source) and linewidth (i.e. $V_{\rm op} \sim -300\pm150$ km~s$^{-1}$) as the 22 GHz H$_{2}$O maser line detection reported in Impellizzeri et al. (2008). Given this assumption, the expected sky frequency distribution of the 380 GHz line ranges from 104.49$\sim$104.52 GHz. As a result, we adopted channels with frequencies ranging from 104.14$\sim$104.48 GHz and 104.57$\sim$104.95 GHz as the line-free channels for continuum subtraction. 

Nevertheless, the subsequent imaging with $TCLEAN$ shows that tentative weak line features can be seen between 104.20 GHz and 104.75 GHz, and this frequency range is significantly wider than what we assumed. Therefore, instead of using baseline channels as described above, we re-do the continuum subtraction by using channels with frequencies ranging from 104.02$\sim$104.20 GHz and 104.75$\sim$104.95 GHz as the baseline channels. Analysis of the line spectrum resulting from spectral-line imaging with $TCLEAN$ after continuum subtraction ensured that the signals in these channels are dominated by noise, suggesting that these are the right channels for continuum subtraction.

Given the broad spectral line distribution seen in spw 25, to perform reliable continuum subtraction in spw 27 to search for the 336 GHz line, we assumed that the 336 GHz line has the same line center and linewidth of the 380 GHz line and adopt the frequency ranges of 91.95$\sim$92.15 GHz and 92.65$\sim$92.85 GHz for continuum subtraction. These frequency ranges corresponds to similar velocity ranges covered by the baseline channels used for continuum subtraction in spw 25.

After the continuum subtraction, we performed the spectral-line imaging with $TCLEAN$. Since angular resolution is not critical for maser detection, we adopt natural weighting for the spectral line imaging to achieve the highest sensitivity for detection. The resulting size of the synthesis beam is 0$\mydprime$.64$\times$0$\mydprime$.38, with PA of $-$61$^{\circ}$. The 1$\sigma$ noise per 0.97708 MHz channel is 0.45 mJy. Finally, for both continuum and spectral-line imaging, we do not perform primary beam correction for our images because the source size is located near the phase center and is significantly smaller than the primary beam size.

\begin{table*}
\tbl{Measurement Results.}{%
\begin{tabular}{lcrrrrrr}
\hline
Measurement & Unit & A1 & A2 & B & C & G & X \\     
 \hline
RA ~~(continuum)  & arcsec  & 0.0$\pm$0.008 & 0.132 $\pm$0.010 & $-0.583\pm0.001$  &       $-$1.948$\pm$0.003   &  $-$1.055$\pm$0.012 & $-$1.440$\pm$0.016  \\
      DEC (continuum) & arcsec & 0.0$\pm$0.005  & 0.403$\pm$0.006  & $1.933\pm0.001$ &  0.303$\pm$0.002  &  ~0.656$\pm$0.009    &  2.113$\pm$0.012   \\
RA ~ (continuum)    & arcsec  & 0.583$\pm$0.008 & 0.715$\pm$0.010 & 0$\pm$0.001  &       $-$1.365$\pm$0.003   &  $-$0.472$\pm$0.012 & $-$0.857$\pm$0.016   \\
DEC (continuum) &  arcsec & $-$1.933$\pm$0.005  & $-$1.530$\pm$0.006  & 0$\pm$0.001 &  $-$1.630$\pm$0.002  &  $-$1.277$\pm$0.009    &   0.180$\pm$0.012  \\
RA ~~(380 GHz line)    & arcsec  & 0.594$\pm$0.031 & 0.716$\pm$0.040 & 0$\pm$0.001  &  ---   &  --- & ---   \\
DEC (380 GHz line) &  arcsec & $-$1.878$\pm$0.020  & $-$1.529$\pm$0.025  & 0$\pm$0.001 &  ---  &  ---   &  ---  \\
f$_{92}$    & mJy   &  13.43$\pm$0.90  & 11.50$\pm$0.83  &  5.15$\pm$0.03  &  1.94$\pm$0.02    & ---  &  ---   \\ 
f$_{93}$    & mJy   &  12.99$\pm$0.88  & 11.23$\pm$0.78 &  5.00$\pm$0.03  &  1.92$\pm$0.03    &  ---  &   ---   \\ 
f$_{104}$  & mJy   &  11.18$\pm$0.61    & 9.52$\pm$0.48  &  4.23$\pm$0.04  &  1.75$\pm$0.03      & ---  &   --- \\
f$_{106}$ & mJy & 10.94$\pm$0.52 & 9.25$\pm$0.51  &  $4.18\pm0.03$  &  1.59$\pm$0.02   & ---  &  ---     \\ 
f$_{\rm Line}$ & mJy/beam.km/s & 73.9$\pm$5.5 & 87.1$\pm$7.6 & --- & --- & --- & --- \\ 
spectral index & --- & $-$1.46$\pm$0.03 &  $-$1.57$\pm$0.03    & $-$1.53$\pm$0.04 & $-$1.29$\pm$0.27  & ---  & ---  \\      
 \hline
 \end{tabular}}\label{tab:first}
\begin{tabnote}
Column (1): Measurement type. The first two rows show the position measurements relative to the peak of the A1 component. The 3rd and 4th rows show the positions relative to the B component. f$_{92}$, f$_{93}$, f$_{104}$, f$_{106}$ represent the continuum flux density measurement at the (band-center) frequency of 92.400 GHz, 93.981 GHz, 104.488 GHz, and 106.087 GHz, respectively. f$_{\rm Line}$ represents the peak flux of the elliptical gaussians fitted to the A1 and A2 components in the integrated 380 GHz line intensity map; column (2): unit of the measurement; column (3)$-$(6): best-fit results for the A1, A2, B, C component of the lensed QSO. column (7)$-$(8): The positions of the gravitational lenses G and X taken from the CASTLES survey\footnote{https://www.cfa.harvard.edu/castles/Individual/MG0414.html}.
\end{tabnote}
\end{table*}

\begin{figure*}[ht] 
\begin{center} 
\vspace*{0 cm} 
\hspace*{-1 cm} 
\includegraphics[angle=0, scale=0.7]{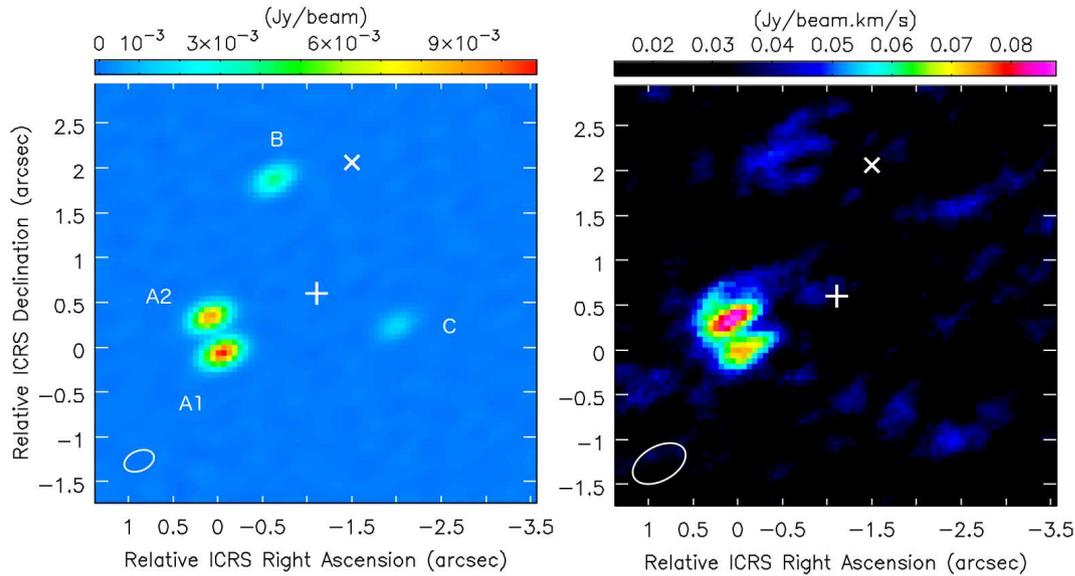}
  \vspace*{0.0 cm} 
\caption{\label{fig:image} {\bf Left panel:} The continuum emission from MG J0104+0534 at 106.081 GHz.  In the figure, A1, A2, B, and C label the four lensed spots of the QSO. The white plus and cross sign show the position of the primary (G) and secondary (X) foreground lens galaxies (Ros et al. 2000), respectively. Their positions are shown in the last two columns in Table 1.  {\bf Right panel:} The integrated-intensity map of the 380 GHz line made by integrating the data cube from 104.2 GHz to 104.75 GHz along the frequency axis.  }
\end{center} 
\end{figure*}

\section{Results}
\subsection{Continuum Emissions}
The continuum image of MG J0414+0534 obtained at 106.87 GHz (see the left panel of Figure 1) shows four unresolved lensed spots A1, A2, B, and C. To obtain precise positions of these spots for lens modeling, we fit elliptical Gaussians to these spots. The measured positions relative to A1 (and B) are shown in Table 1. These positions are consistent with those measured at 22 GHz by Katz, Moore, \& Hewitt (1997) with the VLA. 

In Table 1, we also show the measurements of flux densities of the four lensed spots at all four spectral windows. Assuming a spectral model of the form $f_{\nu}$ = $f_{0}$$\nu^{\alpha}$, the spectral indices $\alpha$ of all lensed spots are consistent with $\alpha$ $\sim$ $-$1.5. Averaging these measurements gives a best estimate of $\alpha$ $=$ $-$1.51$\pm$0.02. This value is consistent with the spectral index distribution of radio jets of quasars at centimeter wavelengths (e.g. Hovatta et al. 2014), suggesting that the observed continuum emissions may come from radio jets. However, we note that the observed spectral index is relatively steeper in comparison with the typical value (i.e. $\alpha$ $=$ $-$1.1) for quasars. If the observed emissions indeed come from jet, it is likely that the spectral index steepening result from the radiative-cooling effect (e.g. Carilli \& Barthel 1996). This is because the synchrotron cooling timescale of electrons is $t_{\rm cool}$ $\propto$ $\nu^{-1/2}$ (Piran 2005) and the rest-frame continuum emission in our observation is at $\sim$380~GHz. Therefore, the cooling timescale of the electrons responsible for the jet emission seen in our ALMA observation must be significantly shorter than that for the jet seen in centimeter wavelengths, and this could lead to the observed steep spectral index. 

Alternatively, the steep spectral index at the rest frame frequency of $\sim$380 GHz could also be explained if the observed radio emission come from supernova remnants (e.g. Indebetouw et al. 2014). However, this explanation will require the star formation rate (SFR) to be extremely high if one attributes all of the observed radio emission to the supernovae (SNe). To give a lower limit of the required SFR to explain the observed radio emission, we assume that all of the SNe that account for the observed radio emission happen to attend the peak radio luminosities after their explosions and the peak spectral radio luminosity of each individual supernova (SN) ($L_{\nu,p}$) has the extreme value of 10$^{29}$ erg~s$^{-1}$~Hz$^{-1}$ at the observing frequency of 5 GHz (Kamble et al. 2016). Given the spectral index of $\alpha$ $=$ $-$1.5, the peak spectral radio luminosity of a SN at 380 GHz can be inferred to be $L_{\nu,p}$(SN) $=$ 1.5$\times$10$^{26}$ erg~s$^{-1}$~Hz$^{-1}$. 

Using the luminosity distance of 22197 Mpc for MG J0414$+$0534 and a magnification factor of 13.1 for the A1 component (see the supplementary information provided by Implellizzeri et al. 2008), the (unlensed) spectral radio luminosity\footnote{This luminosity is estimated using the observed flux density (i.e. 11.18 mJy) for the A1 component at 104 GHz} of the QSO at the rest frame frequency of $\sim$380 GHz is $L_{\nu}$(QSO)$\sim$5$\times$10$^{33}$ erg~s$^{-1}$~Hz$^{-1}$. From this value, one can infer that the number of SNe required for explaining this radio luminosity is $N_{SNe}$ $=$ $L_{\nu}$(QSO)/$L_{\nu,p}$(SN) $\sim$3$\times$10$^{7}$. 

By making an extreme assumption that these SNe maintain the peak luminosity for $\sim$1000 days, this would imply that the rate for these SNe (which happen to attend their peak radio luminosity at the time of our observation) to occur is $R_{SNe}$ $\sim$1$\times$10$^{7}$ yr$^{-1}$. Using the relationship between $R_{SNe}$ and SFR of $R_{SNe}$ $\sim$(0.01/$M_{\odot}$)$\times$SFR (Hopkins \& Beacom 2006; Botticella et al. 2012), the inferred lower limit of the starformation rate is $\sim$10$^{9}$ $M_{\odot}$~yr$^{-1}$. This value is not only more than five orders of magnitude higher than the highest SFR observed in the high-z Universe (e.g.  Barger et al. 2014), but also implies that the starformation will be totally quenched in MG J0414$+$0534 in a timescale significantly shorter than the age of the galaxy. Given that a substantial amount of molecular gas is still available in MG J0414$+$0534 (Stacy \& McKean 2018), it is not likely that the starformation in this galaxy has extinguished. Therefore, we can conclude that SNe cannot account for the majority of the observed radio emission from MG J0414$+$0534 and provide a satisfactory explanation for the observed steep spectral index.

%flux ratios : compare the A1/A2 flux ratio between optical, IR, and submm. Are the ratios consistent ?

% The position uncertainties shown here include both the measurement uncertainties from CASTLE and the systematic position offsets between CASTLE and measurements from Katz, Moore, \& Hewitt (1997), which well match the values from our ALMA observation.

\begin{figure*}[ht] 
\begin{center} 
\vspace*{0 cm} 
\hspace*{-1 cm} 
\includegraphics[angle=0, scale=0.6]{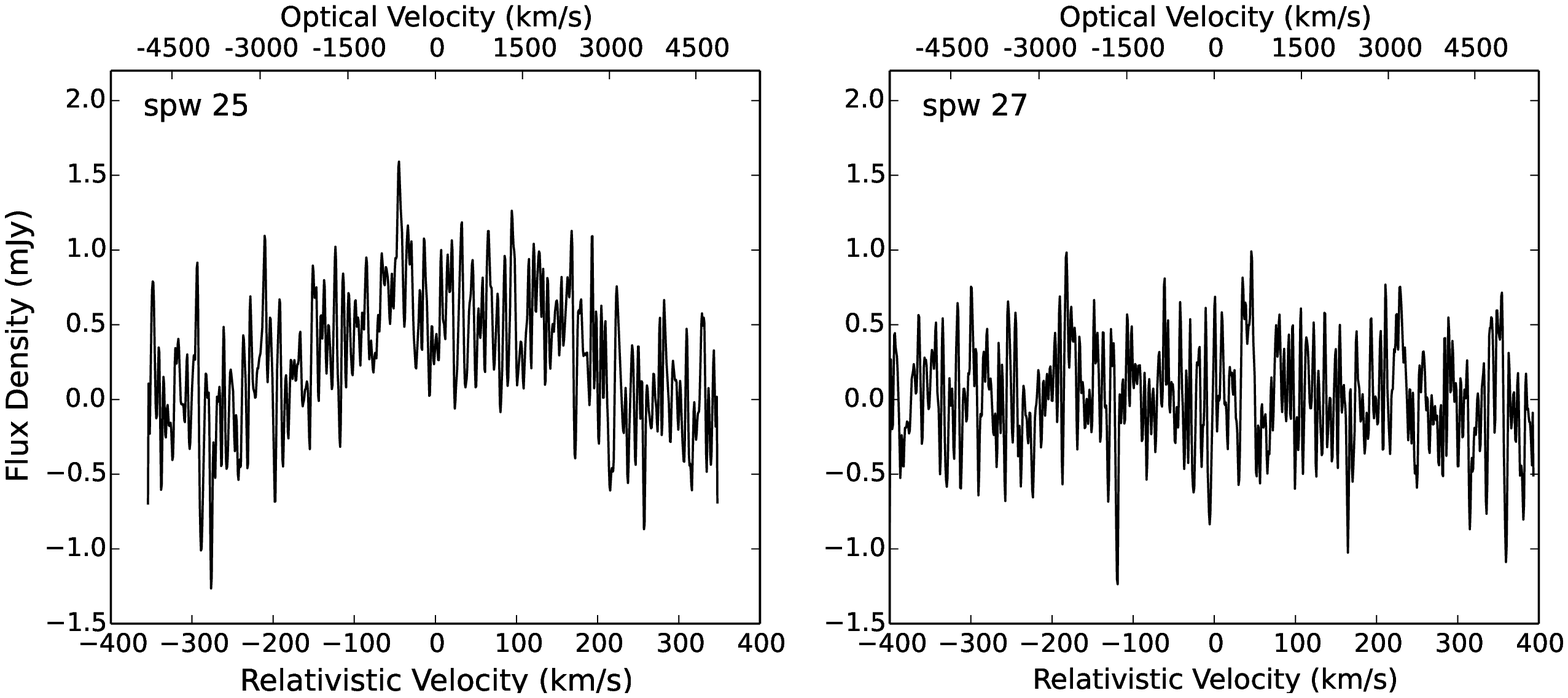}
\vspace*{0.0 cm} 
\caption{\label{fig:spec} {\bf Left Pannel} : The 380 GHz line spectrum of MG J0104+0534 made from summing all emissions from the A1 and A2 component seen in Figure 1.     {\bf Right panel:} The 336 GHz line spectrum of the QSO. The upper and lower horizontal axes show the velocity of each spectral channel in terms of the optical convention and the relativistic velocity, respectively. The optical velocities are shown here for easier comparison between our spectrum and the 22 GHz maser spectrum measured by Impezzelleri et al. (2008).}
\end{center} 
\end{figure*} 

\subsection{A Tentative Detection of the 380 GHz Line}
\subsubsection{ The Integrated Intensity Map}
To identify the 380 GHz line from our data, we first measure the line spectrum by summing all emissions in each channel of the data cube for spw 25 from the A1 and A2 component. Careful examination of the spectrum suggests that weak line features can be seen between 104.2 GHz and 104.75 GHz in the spectrum with the signal-to-noise ratio of $\sim$2$-$3 $\sigma$. To examine whether these features are simply noise, we integrate the data cube (with continuum removed) along the frequency axis between 104.2 GHz and 104.75 GHz and make the integrated intensity (MOM0) map (see the right panel of Figure 1). The line emissions can be seen at the A1 and A2 positions, suggesting that the tentative line emissions we see might be real signals. 

This image shows two unexpected features. First, the flux ratio between the A1 and A2 line component (A2/A1 $=$ 1.2) is different from what we see in the continuum image  (A2/A1 $=$ 0.8). This is unexpected if the line emissions from A1 and A2 receive the same amount of magnification as the continuum emissions. Second, the lensed line emission at the A2 component is slightly resolved and show a weak extended structure pointing toward north. If the extended emissions we see are confirmed with future high sensitivity observation, this would suggest that the line emissions from A2 may not be entirely associated with continuum emission from the QSO as expected, and some of the emission could be associated with more extended structures such as molecular outflows or starforming activity. 

\subsubsection{The Line Spectra}
In the left panel of Figure 2, we show the 380 GHz line spectrum taken at spw 25. This spectrum was made from summing all line emissions in the data cube from the A1 and A2 positions that can be directly seen in the right panel of Figure 1. The 1-$\sigma$ noise in the spectra after Hanning smoothing over 7 channels is 0.30 mJy.

The spectrum reveals a broad spectral distribution. By fitting a Gaussian profile to the spectrum, we obtained a full width at half maximum (FWHM) of $\sim$290 km~s$^{-1}$  (with relativistic correction). This linewidth is significantly broader than the 22 GHz maser line detected in Castangia et al. (2011), which reported a maser linewidth of 174 km~s$^{-1}$ for the dominant line feature in terms of the optical velocity convention (i.e. $\sim$13 km~s$^{-1}$ after relativistic correction). Based on Equation (1) in Castangia et al. (2011), with the line luminosity multiplied by a factor of 13.9 to account for relativistic correction for the velocity width and a magnification factor of 27.9 (i.e. the total lensing magnification for A1 and A2; see the supplementary information provided by Implellizzeri et al. 2008), we get a total isotropic (unlensed) luminosity $L_{\rm H_{2}O}$(380 GHz) of $\sim$5400000 $L_{\odot}$, which is brighter than any 22 GHz masers that have ever been discovered (Barvainis \& Antonucci 2005;  Impellizzeri et al. 2008). If all of 380 GHz line emissions come from maser emissions, MG J0414$+$0534 will become the first example of a ``gigamaser" galaxy $-$ a galaxy with maser luminosity billion times higher than galactic masers (Lo 2005). Nonetheless, we note that, given the broad spectral distribution, we cannot fully exclude the possibility that at least part of the 380 GHz line emissions are thermal emissions coming from starforming regions in the QSO host galaxy. As we will see in section 4, this is a possibility that is consistent with the results of our gravitational lens modeling, which suggests that some of the 380 GHz line emissions may come from starforming regions in the host galaxy.

\subsection{The Non-Detection of the 336 GHz Line}
To search for  the 336 GHz line, we performed the same procedure as describe in section 2 and section 3.2.1 for the data cube imaged at spw 27. Unfortunately, no detection can be see in the spectrum (see the right panel of Figure 2) and we can place a 3-$\sigma$ upper limit of the 336 GHz line to be 1.0 mJy.

\section{Discussion}
The 380 GHz maser line is one of the luminous submillimeter maser transitions (Gray et al. 2016) that cannot be directly accessible by currently available radio telescopes for sources in the local Universe because of significant atmospheric absorption near the rest frequencies of these maser lines. However, thanks to the high redshift nature of our target source, the 380 GHz maser transition gets redshifted to a spectral window in which high sensitivity spectral line observations with ALMA become possible.

In our present work, our main goal is to conduct the first search of the 380 GHz maser line from a high redshift source and explore the possibility of using the line as a new tool to determine accurate black hole masses in high redshift AGNs. Despite of the fact that we do not detect strong enough line emissions for future VLBI follow-ups, the tentative detection of the 380 GHz shown in section 3 would worth future high sensitivity observation with ALMA for its confirmation. This is because if our tentative detection can be confirmed, it may not only demonstrate the first example of a ``gigamaser" system if the majority of the line emissions are maser in nature, but also provide a firm basis for future search of the 380 GHz line from high-z Type-2 lensed quasars/AGNs, from which maser emissions are expected to be significantly stronger than Type-1 AGN such as our target source MG J0414$+$0534 (Braatz et al. 1997; Kuo et al. 2018). In addition, it will be the first detection of the 380 GHz water emissions from astronomical sources.

Given that the full spectral width of the line profile appears to be comparable to the velocity width of the spectral window ($\sim$700 km~s$^{-1}$), one could question that whether what wee see here are actually residual continuum emissions as a result of the residual bandpass. To argue against this possibility, we note that if the broad spectral profile is simply residual continuum, the 380 GHz A1/A2 line flux ratio should be similar to the ratio derived from the continuum image, but this is not the case. In addition, by examining the bandpass solutions for different antennas from the calibration process, we do not see a characteristic bandpass shape as seen in Figure 2 and the bandpass functions varies substantially, suggesting that the residual bandpass should be negligible after averaging the entire dataset. Furthermore, since the bandpass corrections were also applied to the phase calibrators, if some bandpass residuals survived after data averaging, we should have seen the same residual in the spectra of the phase calibrators, but no such bandpass residuals can be seen. Given these reasons, we believe that what we see in Figure 2 is likely to be a real detection of the 380 GHz line despite of the low signal-to-noise ratio. Nonetheless, we want to emphasize that our current detection is still tentative. Higher sensitivity observations is definitely needed to confirm our currently tentative detection.

In the following, we will try to explore what we might see in future higher sensitivity observation of the 380 GHz line towards MG J0414$+$0524 under the assumption that our tentative line detection is real. The discussion below may provide guidelines that help designing future observation in a way that may not only confirm the existence of the 380 GHz line, but also provide deeper insights into the nature and origin of the 380 GHz maser emissions.

\subsection{The Origin of the 380 GHz Line}

Given the positional association of the 380 GHz line emission with the continuum from radio jet, the broad spectral profile of the tentative 380 GHz line detection toward MG J0414+0534 is reminiscent of the 22 GHz maser spectrum of the nearby maser galaxy NGC 1052, which also shows a broad spectral distribution spanning over $\sim$400 km~s$^{-1}$ in the frequency space (Kameno et al. 2005). In this nearby maser galaxy which hosts a Seyfert 2/Liner nucleus, a nearly symmetric double-sided radio jet can be seen (e.g. Wrobel 1984; Kellermann et al. 1998). Using the VLBI technique, Claussen et al. (1998) and Sawada-Satoh et al. (2009) show that the 22 GHz H$_{2}$O maser emission is concentrated into two groups whose projected positions lie on top of the knots in the radio jets. Claussen et al. (1998) suggest that the observed maser emission in NGC 1052 could be powered by slow, non-dissociative shocks that are driven into circumnuclear dense molecular cloud. On the other hand, Kameno et al. (2005), Sawada-Satoh et al. (2008), and Sawada-Satoh et al. (2016) proposed a torus scanner model in which the maser emission arises from a heated molecular layer or X-ray dissociation region (XDR; Neufeld et al. 1994; Neufeld \& Maloney 1995) which lies close to the inner surface of the torus at the central region of the active galactic nucleus (AGN) in NGC 1052.

In the case of our target source MG J0414$+$0534, given the prominent radio jet observed by Ros et al. (2000) and Volino et al. (2010) as well as the broad spectral distribution of the 380 GHz H$_{2}$O maser emission seen in our ALMA observation, it is plausible to speculate that the 380 GHz line may have the same physical origin or excitation mechanism as the 22 GHz maser observed in NGC 1052. However, we note that no matter whether the 380 GHz line is powered by jet or arises from a XDR, it is required that the maser emissions must lie very close to the jet or the central AGN. If the above-mentioned mechanisms account for all of the 380 GHz line emission we observed, it would be difficult to explain the A1/A2 line flux ratio anomaly discussed in section 3.2.1 because the line and continuum emission coming from the same position in the sky should receive the same amount of magnification. To get a hint for the cause of the unexpected flux ratio and to predict what we would see in future higher sensitivity ALMA observation for detection confirmation, it would be helpful to perform lens modeling.

\begin{figure*}[ht] 
\begin{center} 
\vspace*{0 cm} 
\hspace*{-1 cm} 
\includegraphics[angle=0, scale=0.6]{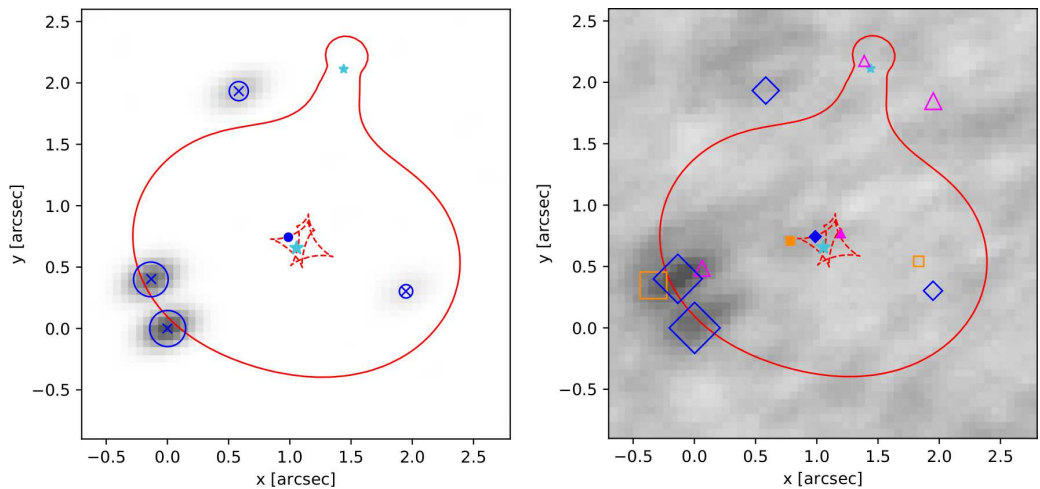}
\vspace*{0.0 cm} 
\caption{\label{fig:lensmodel} {\bf Left panel:} Best-fit lens model of the continuum observations overlaid in grey.  Crosses are observed image positions, filled circle is the modeled source position, open circles are the model predicted image positions with areas proportional to the image magnifications, and the stars indicate the centroids of lens galaxies G (big star) and X (small star).  The solid (dashed) red curve is the critical curve (caustics).  The model fits to the observed positions and fluxes very well with a reduced $\chi^2$ of 1.0. {\bf Right panel:} Hypothetical maser clumps in solid symbols and their corresponding multiple lensed images in open symbols, overlaid on the integrated-intensity map of the 380 GHz line in grey. These maser clumps displaced from the continuum position produce lensed image only in A1 but not in A2, and could explain the brighter A2 line image in comparison to A1. }
\end{center} 
\end{figure*}

\subsection{Lens Modeling}

We model the lens system with {\sc Glee}, a lensing software developed by A.~Halkola and S.~H.~Suyu \citep{SuyuHalkola10, SuyuEtal12}.  As previously noted by \citet{RosEtal00} and also \citet{MacLeodEtal13}, there is an object X near image B, in addition to the primary lens galaxy G.  We use a singular isothermal elliptical (SIE) mass distribution \citep{KassiolaKovner93} to model the primary lens galaxy G with 5 parameters (centroid $x_{\rm G}$, $y_{\rm G}$, axis ratio $q_{\rm G}$, position angle $\phi_{\rm G}$ and Einstein radius $\theta_{\rm E,G}$), and a singular isothermal sphere (SIS) for lensing object X with 3 parameters (centroid $x_{\rm X}$, $y_{\rm X}$, and Einstein radius $\theta_{\rm E,X}$).  We adopt a prior on the centroids of G and X based on the observed light distribution (listed in Table 1), and uniform prior on the other parameters.  In addition to G and X, we include a constant external shear component with shear strength $\gamma_{\rm ext}$ and position angle $\phi_{\rm ext}$ to account for the shear from the lens environment.

To constrain these model parameters, we use the four observed image positions of the QSO (A1, A2, B and C) and the fluxes of A1 and A2 from the 106 GHz continuum observations listed in Table 1.  Our model of SIE(G)+SIS(X)+shear for the system fits well to the observed images and fluxes, with a reduced $\chi^2$ of 1.0. In the left panel of Figure 3, we show our best-fit lens model overlaid on the continuum image in grey shades: the crosses are the observed image positions, open circles are the predicted positions with the enclosed area in the circle proportional to the model predicted magnification, filled circle is the source position, solid red curve is the critical curves, dashed red curves are the caustics, the big star marks the centroid of G, and the small star marks the centroid of X.

Most of the lens models constrained by the image positions A1 and A2 from the continuum image yields a magnification of A1 that is close to or slightly larger than A2.  What then causes A2 to be brighter than A1 in the integrated-intensity map of the 380 GHz line shown in the right-hand panel of Figure 1 ?  
%As indicated in Table 1, the positions of A1 and A2 of the maser emission are different from those of the continuum, albeit with larger uncertainty.  %We therefore explore the multiple image configurations of a strongly lensed maser clump that is located at a different position as the continuum emission region on the source plane.  

To explain the flux ratio anomaly, we consider a multi-component maser model. The right-hand panel of Figure 3 shows three hypothetical maser clumps. The maser clump that dominates observed line emission is located at the continuum emission position (the filled diamond).  The filled (magenta) triangle and filled (orange) square represent the other two maser clumps (possibly associated with starformation) that are spatially displaced from continuum position by $\sim$1.5 kpc on the source plane. The open diamond, square and triangle are the lensed images of the corresponding filled symbols on  the image plane. The area enclosed by the open symbol is proportional to the magnification factor.  

This figure shows that while the dominant maser clump at the continuum position produces lensed images in both A1 and A2, the maser clumps represented by the filled triangle and square would only produce observable lensed images in A2, but not in A1. As a result of these additional lensed images, the fluxes add up in A2 and lead to brighter A2 line image in comparison with A1. 

If our simple model is correct, it not only explains the flux ratio anomaly, but also explains why the A2 line image is slightly resolved.
Moreover, it demonstrates that the maser emissions are not only associated with the AGN, but may also come from starforming regions in the QSO host galaxy. 

To test our simple model, future ALMA observation needs to improve the angular resolution by at least a factor of 2 in comparison with our present observation in order to resolve the lensed maser emissions coming from different maser clumps in the source plane. In addition, the sensitivity of the observation also needs to improve by a least a factor of 2$-$3 in order to detecting the other multiple lensed images from the displaced maser clumps.

\section{Conclusion}
In our newly launched program of searching for 380 GHz H$_{2}$O maser emissions from high-z lensed galaxies to explore whether the 380 GHz maser line can be used as a new tool to study the vicinity of supermassive black holes in the high-z Universe, our ALMA observation on the high-z lensed QSO MG J0414$+$0534 at z $=$ 2.639 gives the first hint of the existence of the 380 GHz maser line. Although our detection is tentative, our observation does show that detecting the 380 GHz maser line from high-z lensed galaxies is possible.

Because of the broad spectral profile of the tentative maser detection, in order to confirm the 380 GHz maser detection in future ALMA observations on MG J0414$+$0534, one must arrange the spectral windows in such a way that the total bandwidth that covers the line has to be substantially broader than 1 GHz so that one can see full spectral profile and its baseline more clearly and exclude the possibility of false line features caused by imperfect continuum subtraction. In addition, our simple lens model that is designed to explain the maser flux ratio anomaly suggests that the 380 GHz line emissions come from two or three spatially displaced H$_{2}$O maser components in the QSO, with the dominant one located at the continuum emission position and the other one(s) displaced from the continuum by $\sim$1.5 kpc on the source plane. 

Therefore, in addition to requiring longer integration time to confirming the line detection with higher signal-to-noise ratio, achieving higher angular resolution that resolves the maser clumps around A2 and sufficient sensitivity for detecting the other multiple lensed images would not only help decipher the origin of the flux anomaly and but also help revealing the origin of the 380 GHz line emissions in this high-z gravitationally lensed system.

\section*{Acknowledgement} 
We thank the anonymous referee for providing highly valuable comments for helping improving this manuscript. We also thank Crystal Brogan at NRAO for her valuable comments on our ALMA data reduction. We thank S.~Schuldt for sharing his python plotting script. SHS thanks the Max Planck Society for support through the Max Planck Research Group. This publication is supported by Ministry of Science and Technology, R.O.C. under the project 107-2119-M-110-005. This research has made use of NASA's Astrophysics Data System Bibliographic Services, and the NASA/IPAC Extragalactic Database (NED) which is operated by the Jet Propulsion Laboratory, California Institute of Technology, under contract with the National Aeronautics and Space Administration. This paper makes use of the following ALMA data: ADS/JAO.ALMA\#2017.0.00316.S. ALMA is a partnership of ESO (representing its member states), NSF (USA) and NINS (Japan), together with NRC (Canada), MOST and ASIAA (Taiwan), and KASI (Republic of Korea), in cooperation with the Republic of Chile. The Joint ALMA Observatory is operated by ESO, AUI/NRAO and NAOJ.


\begin{thebibliography}{}       
\bibitem[Barger et al. (2014)]{barger14} Barger, A. J., Cowie, L. L., Chen, C.-C., Owen, F. N., Wang, W.-H., Casey, C. M., Lee, N., Sanders, D. B., Williams, J. P. 2014, ApJ, 784, 9
\bibitem[Bernnert et al. (2009)]{ber09} Bennert, N., Barvainis, R., Henkel, C., Antonucci, R. 2009, ApJ, 695, 	276
\bibitem[Braatz et al. (1997)]{bra97} Braatz, J. A., Wilson, A. S., Henkel, C. 1997, ApJS, 110, 321
\bibitem[Braatz et al. (2018)]{bra18}   Braatz, J., Condon, J., Henkel, C. et al. 2018, IAUS, 336, 86
\bibitem[Botticella et al. (2012)]{bol12} Botticella, M. T., Smartt, S. J., Kennicutt, R. C., Cappellaro, E., Sereno, M., Lee, J. C. 2012, A\&A, 537, 132
\bibitem[Castangia et al. (2011)]{cas11}  Castangia, P., Impellizzeri, C. M. V., McKean, J. P., Henkel, C., Brunthaler, A., Roy, A. L., Wucknitz, O., Ott, J., Momjian, E. 2011, A\&A, 529, 150
\bibitem[Carilli \& Barthel (1996)]{cb96} Carilli, C. L., Barthel, P. D. 1996, ARA\&A, 7, 1
\bibitem[Claussen et al. (1998)]{clau98}  Claussen, M. J., Diamond, P. J., Braatz, J. A., Wilson, A. S., Henkel, C. 1998, ApJ,  500, 129         
\bibitem[de Jong (1973)]{dejong73} de Jong 1973, A\&A, 26, 297
\bibitem[Gao et al. (2016)]{gao16} 
Gao, F., Braatz, J. A., Reid, M. J., Lo, K. Y., Condon, J. J., Henkel,
C., Kuo, C. Y., Impellizzeri, C. M. V., Pesce, D. W., Zhao, W. 2016,
ApJ, 817, 128                                                              
\bibitem[Gray et al. (2016)]{gray16} Gray, M. D., Baudry, A., Richards,
A. M. S., Humphreys, E. M. L., Sobolev, A. M.; Yates, J. A. 2016,
MNRAS, 456, 347    
\bibitem[Greenhill et al. (2003)]{ghl03}  Greenhill, L. J., Booth, R. S., Ellingsen, S. P., Herrnstein, J. R., Jauncey, D. L., McCulloch, P. M., Moran, J. M., Norris, R. P., Reynolds, J. E., Tzioumis, A. K. 2003, ApJ, 590, 162
\bibitem[Herrnstein et al. (1999)]{hern99}  Herrnstein, J. R., Moran, J. M., Greenhill, L. J., Diamond, P. J., Inoue, M., Nakai, N., Miyoshi, M., Henkel, C., Riess, A. 1999, Nature, 400, 539
\bibitem[Hopkins \& Beacom (2006)]{hb06} Hopkins, A. M., Beacom, J. F. 2006, ApJ, 651, 142
\bibitem[Hovatta et al. (2014)]{hov14} Hovatta, T., Aller, M. F., Aller, H. D., Clausen-Brown, E., Homan, D. C., Kovalev, Yu. Y., Lister, M. L., Pushkarev, A. B., Savolainen, T. 2014, AJ, 147, 143        
\bibitem[Impellizzeri et al. (2008)]{vio08} Impellizzeri, C. M. V., McKean, J. P., Castangia, P., Roy, A. L., Henkel, C., Brunthaler, A., Wucknitz, O. 2008, Nature, 456, 927
\bibitem[Indebetouw et al. (2014)]{remy14}   Indebetouw, R., Matsuura, M., Dwek, E et al. 2014, ApJ, 782, 2
\bibitem[Kamble et al. (2016)]{kam16} Kamble, A., Margutti, R., Soderberg, A. M.. Chakraborti, S., Fransson, C., Chevalier, R., Powell, D., Milisavljevic, D., Parrent, J., Bietenholz, M. 2016, ApJ, 818, 111
\bibitem[Kameno et al. (2005)]{kam05} Kameno, S., Nakai, N., Sawada-Satoh, S., Sato, N., Haba, A. 2005, ApJ, 620, 145
\bibitem[Kassiola \& Kovner (1993)]{KassiolaKovner93} 
Kassiola, A., Kovner, I. 1993, ApJ, 417, 450
\bibitem[Katz, Moore, \& Hewitt (1997)]{kmh97} Katz, C. A., Moore, C. B., Hewitt, J. N. 1997, ApJ, 475, 512
\bibitem[K\ddot{o}nig et al. (2017)]{konig17} K${\rm \ddot{o} }$nig, S., Martin, S., Muller, S. et al. 2017, A\&A, 602, 42
\bibitem[Kellermann et al. (1998)]{kelman98} Kellermann, K. I., Vermeulen, R. C., Zensus, J. A., \& Cohen, M. H. 1998, AJ, 115, 1295
\bibitem[Kuo et al. (2011)]{kuo11} 
Kuo, C. Y.; Braatz, J. A.; Condon, J. J.; Impellizzeri, C. M. V.;
Lo, K. Y.; Zaw, I.; Schenker, M.; Henkel, C.; Reid, M. J.; Greene,
J. E. 2011, ApJ, 727, 20     
\bibitem[Kuo et al. (2013)]{kuo13}  Kuo, C. Y., Braatz, J. A., Reid, M. J., Lo, K. Y., Condon, J. J., Impellizzeri, C. M. V., Henkel, C. 2013, ApJ, 767, 155
\bibitem[Kuo et al. (2015)]{kuo15} Kuo, C. Y., Braatz, J. A., Lo, K. Y., Reid, M. J., Suyu, S. H., Pesce, D. W., Condon, J. J., Henkel, C., Impellizzeri, C. M. V. 2015, ApJ, 800, 26
\bibitem[Kuo et al. (2018)]{kuo18}	Kuo, C. Y., Constantin, A., Braatz, J. A., Chung, H. H., Witherspoon, C. A., Pesce, D., Impellizzeri, C. M. V., Gao, F., Hao, Lei, Woo, J.$-$H., Zaw, Ingyin 2018, ApJ, 860, 169
\bibitem[Lo (2005)]{lo05} 	Lo, K. Y. 2005, ARA\&A, 43, 625
\bibitem[McKean et al. (2011)]{mck11} McKean, J. P., Impellizzeri, C. M. V., Roy, A. L., Castangia, P., Samuel, F., Brunthaler, A., Henkel, C., Wucknitz, O. 2011, MNRAS, 410, 2506
\bibitem[MacLeod et al.~(2013)]{MacLeodEtal13} MacLeod, C. L., Jones, R., Agol, E., Kochanek, C. S. 2013, ApJ, 773, 35
\bibitem[McMullin et al. (2007)]{mcm07} McMullin, J. P., Waters, B., Schiebel, D., Young, W., \& Golap, K. 2007, Astronomical Data Analysis Software and Systems XVI (ASP Conf. Ser. 376), ed. R. A. Shaw, F. Hill, \& D. J. Bell (San Francisco, CA: ASP), 127
\bibitem[Neufeld et al. (1994)]{neu94}  Neufeld, D. A., Maloney, P. R., Conger, S. 1994, ApJ, 436, 127	
\bibitem[Neufeld \& Maloney (1995)]{nm95} Neufeld, D. A., Maloney, P. R. 1995, ApJ, 447, 17
\bibitem[Piran, T. (2005)]{piran05} Piran, T. 2005, Rev. Mod. Phys., 76, 1143
\bibitem[Ros et al. (2000)]{RosEtal00}
Ros, E., Guirado, J. C., Marcaide, J. M., P�rez-Torres, M. A., Falco, E. E., Mu�oz, J. A., Alberdi, A., Lara, L. 2000, \aap, 362, 845
\bibitem[Reid et al. (2009)]{reid09}  Reid, M. J., Braatz, J. A., Condon, J. J., Greenhill, L. J., Henkel, C., Lo, K. Y. 2009, ApJ, 695, 287
\bibitem[Sawada-Satoh et al. (2008)]{ss08}   Sawada-Satoh, S., Kameno, S., Nakamura, K. 2008, ApJ, 680, 191
\bibitem[Sawada-Satoh et al. (2009)]{ss09} Sawada-Satoh, S., Kameno, S., Nakamura, K., Namikawa, D., Shibata, K. M. 2009, AN, 330, 141
\bibitem[Sawada-Satoh et al. (2016)]{ss16} Sawada-Satoh, S., Roh, D.-G., Oh, S.-J. et al. 2016, ApJ, 830, 3
\bibitem[Shen 2013]{shen13}  Shen, Y. 2013, BASI, 41, 61
\bibitem[{{Suyu} \& {Halkola}(2010)}]{SuyuHalkola10}
{Suyu}, S.~H., \& {Halkola}, A. 2010, \aap, 524, A94
\bibitem[Stacy \& McKean (2018)]{sm18}  Stacey, H. R., McKean, J. P. 2018, MNRAS, 481, 40 
\bibitem[{{Suyu} {et~al.}(2012){Suyu}, {Hensel}, {McKean}, {Fassnacht}, {Treu},
  {Halkola}, {Norbury}, {Jackson}, {Schneider}, {Thompson}, {Auger},
  {Koopmans}, \& {Matthews}}]{SuyuEtal12}
{Suyu}, S.~H., {et~al.} 2012, \apj, 750, 10
\bibitem[Volino et al. (2010)]{volino10} Volino, F., Wucknitz, O., Porcas, R. W et al. 2010, Proceedings of the 10th European VLBI Network Symposium and EVN Users Meeting: VLBI and the new generation of radio arrays, 29
\bibitem[Wrobel (1984)]{wrobel84} Wrobel, J. M. 1984, ApJ, 284, 531
\bibitem[Zhao et al. (2018)]{zhao18}   Zhao, W., Braatz, J. A., Condon, J. J., Lo, K. Y., Reid, M. J., Henkel, C., Pesce, D. W., Greene, J. E., Gao, F., Kuo, C. Y., Impellizzeri, C. M. V. 2018, ApJ, 854, 124


\end{thebibliography}
\end{document}